\documentclass[aps,prd,twocolumn,showpacs,floatfix,superscriptaddress,preprintnumbers]{revtex4}
\usepackage{amssymb}
\usepackage{epsfig}
\usepackage{graphicx}
\usepackage{stmaryrd}

\def\be{\begin{equation}}
\def\ee{\end{equation}}
\def\bea{\begin{eqnarray}}
\def\eea{\end{eqnarray}}
\newcommand{\ba}{\begin{array}{c}}
\newcommand{\baz}{\begin{array}{cc}}
\newcommand{\bad}{\begin{array}{ccc}}
\newcommand{\ea}{\end{array}}
\newcommand{\nn}{\nonumber}

\relax

\begin{document}
\preprint{DFPD-09/TH/06}

\title{A dynamical approach to link low energy phases with leptogenesis}
\author{Yin, Lin}
 \email{yinlin@pd.infn.it}
 \affiliation{%
Dipartimento di Fisica `G.~Galilei', Universit\`a di Padova 
\\ 
INFN, Sezione di Padova, Via Marzolo~8, I-35131 Padua, Italy
}%

\date{\today}

\begin{abstract}

If lepton masses and mixings are explained by a flavour symmetry in seesaw model
which leads to $U_{e3}=0$ at leading order, we find that, under reasonable assumptions,
a future observation of low energy leptonic CP violation implies, 
barring accidental cancellations, a lepton asymmetry both in flavoured leptogenesis
and in its one-flavour approximation.
We explicitly implement this approach with a predictive seesaw model for Tri-Bimaximal 
Mixing (TBM) and show how cosmological baryon asymmetry can be 
directly trigged by low energy phases appearing in $U_{e3}$. 
Thanks to this direct correlation we can derive a lower bound on the
reactor angle $\theta_{13}$: $\sin^2 \theta_{13} \gtrsim 0.005$.

\end{abstract}


\maketitle


{\it{A.~Introduction.}}~
All experimental data widely confirm the existence of neutrino masses,
significantly smaller than those of charged fermions. 
Leptonic mixing pattern is also very different from $V_{\rm{CKM}}$ 
because it contains a nearly maximal 
atmospheric angle $\theta_{23}$ and 
a very precise tri-maximally mixed solar angle, 
i.e. $\sin^2 \theta_{12}=1/3$ \cite{TBM0, data}. 
The appealing feature of neutrino mass structure 
can be nicely explained by (type I) seesaw mechanism implemented with
an appropriate flavour symmetry. The super-heavy right-handed Majorana neutrinos, 
added to the standard model (SM), violate the lepton number by two units.
Then their out-of-equilibrium decay can play a role in the generation 
of the observed baryon asymmetry of the universe (BAU) \cite{BAU} through leptogenesis \cite{lepto}. 
CP violation in the leptonic sector is a necessary ingredient 
in order to implement leptogenesis.
However, the seesaw lagrangian generically contains 
a larger number of free parameters than its effective light neutrino sector
and consequently has a poor predictive power.
In particular, in addition to the CP violating phases present in the neutrino
mixing matrix $U_{\rm{PMNS}}$, there are high energy phases not directly observable in
low energy experiments.
Establishing a direct connection between leptogenesis 
and low energy phases is a very important issue and might offer
a possible test of seesaw mechanism.
In one-flavour limit, only high energy phases explicitly 
appear in the lepton asymmetry and such a connection 
is generically believed to arise only from minimal \cite{minimal} /texture 
zeros \cite{zeros} seesaw approaches.
Otherwise, one can take into account flavour effects
and in this case the lepton asymmetries depend on $U_{\rm{PMNS}}$.
An explicit connection between leptogenesis and low energy phases can then
arise when high energy phases are simply assumed to be absent \cite{flavoured}.

In the present letter, we propose a dynamical approach to
link leptogenesis with low energy CP violating phases.
We exploit the possibility that an underlying flavour symmetry naturally leads to both
 $U_{e3} =0$ and vanishing lepton asymmetries at leading order.
We show that, under reasonable assumptions, 
leptogenesis is always related with 
the low energy CP violating phases both in ``flavoured'' and ``unflavoured'' regimes. 
As an explicit example, 
we will also present a predictive seesaw model for TBM based on
$A_4$ flavour symmetry in which low energy phases are
responsible for the BAU. 

{\it{B.~General Consideration.}}~
The smallness of the reactor angle $\theta_{13}$ is a popular
ingredient to characterize a flavour symmetry.
We will begin our analysis by recalling the general mass structure 
which automatically leads to a vanishing  $\theta_{13}$ \cite{parity}. 
Consider for a moment the effective lepton 
sector in the flavour basis $-{\cal L}=e^c m'_l e+\nu m'_\nu \nu$
where $m'_l= {\rm {diag}} (m_e, m_{\mu}, m_{\tau})$ and 
\begin{equation} \label{flavourbasis}
m'_{\nu}=U_{\rm{PMNS}}^* D_m \,U^\dagger_{\rm{PMNS}}~
\ee
where $D_m \equiv {\rm {diag}} (m_1, m_2, m_3)$ with $m_i \geq 0$.
Using the standard parametrization
for the neutrino mixing matrix $U_{\rm{PMNS}}$ 
and imposing $U_{e3} = 0$, one can immediately see
that  $m'_{\nu}$ takes the form :
\begin{equation} 
\label{allproposal}
\left( \bad 
a-b &  d \sqrt{2} \cos \gamma &  d  \sqrt{2} \sin \gamma \\[0.2cm] 
 d \sqrt{2} \cos \gamma & a+(c-b)\cos 2\gamma & (c-b) \sin 2\gamma  \\[0.2cm] 
d  \sqrt{2}  \sin \gamma & (c-b) \sin 2\gamma & a+(c-b)\cos 2\gamma \\ 
               \ea   \right)
\end{equation}            
where the parameters depend on the low energy observable quantities
and in particular $\gamma = \theta_{23}$.
This matrix is invariant under
a parity operator $G_{23} \simeq Z_2$:
\be
G_{23}= \left(
\bad
1&0&0\\
0&\cos \gamma & \sin \gamma \\
0& \sin \gamma & -\cos \gamma
\end{array}
\right) \nn
\ee
which exactly corresponds to the
$\nu_\mu-\nu_\tau$ exchange symmetry when $\gamma= \pi/4$ \cite{mu-tau}.
$L_e - L_\mu - L_\tau$ symmetry \cite{LLL} is the most simple
example corresponding to the case $a = b = c =0$.
Bi-Maximal mixing pattern \cite{BI} can be realized by choosing 
simultaneously $\gamma= \pi/4$ and $c=0$
while TBM \cite{TBM} requires $\gamma= \pi/4$ and $c=d$.
Then our following analysis can in principle work within all these models.
A realistic flavour symmetry group $G_f$ should be usually
 larger than $G_{23}$ in order to enforce also
diagonal and hierarchical charged lepton masses.
Indeed, the peculiar feature of
the flavour basis for this class of models 
can be achieved dynamically by vacuum misalignment
in the spontaneous breaking of $G_f$.
Recently, models in which charged fermion
hierarchies are directly generated by vacuum alignment
that maximally breaks the $\nu_\mu-\nu_\tau$ symmetry \cite{hierarchy, A4Lin}
become of great interest.

Now we consider the seesaw lagrangian 
$$-{\cal L}=\nu^c Y_\nu l h^u+\nu^c M \nu^c + h.c. $$
which gives rise to light neutrino masses $m_\nu= -v_u^2Y_\nu^T M^{-1}Y_\nu$
after the electroweak (EW) symmetry breaking.
The theory can be further implemented by a flavour symmetry
$G_f$ broken spontaneously by a set of flavon fields $\Phi$.
We require that the scalar potential for $\Phi$ allows two 
different alignments in such a way that the effective neutrino
sector preserves a symmetry containing $G_{23}$
and the charged leptons are diagonal.
For the model building, it is quite natural to ensure that
the leading order lepton mixing matrix is already encoded 
 in the right-handed neutrino mass matrix $M$
which should be then diagonalized by the transformation:
\begin{equation}
U_0^\dagger M U^*_0 ={\tt diag}(M_1, M_2, M_3) \equiv D_M~,
\label{U0}
\end{equation}
where  $(U_0)_{13}=0$, $M_i > 0 $ and $U_0$ differs from $U_{\rm{PMNS}}$ by
subleading order contributions $\sim \langle \Phi \rangle /\Lambda \ll 1$.
Observe that any symmetry spontaneously broken by $\Phi$ which can
lead to a mass pattern of the form (\ref{allproposal}) at effective level can be realized 
also for right-handed neutrinos due to their Majorana nature.
Then at leading order we have:
\bea
 m_{\rm{diag}}=v^2 U_0^TY_\nu^TU^+_0M^{-1}_{\rm{diag}}
U_0^\dagger Y_\nu U_0~. \nn
\eea
Defining $\tilde{Y}_\nu = U_0^\dagger Y_\nu U_0$, the solution to the previous
equation is given by
\be
\tilde{Y}_\nu = {\rm{diag}}(\pm \sqrt{m_1M_1},\pm \sqrt{m_2M_2},\pm \sqrt{m_3M_3})/v
\label{yukt}
\ee
and this is equivalent to require that $Y_\nu$ has the same structure
 of M displayed in (\ref{allproposal}).

We suppose that the flavour symmetry is broken at a very high scale 
$10^{12}\, \rm{GeV} < \langle \Phi \rangle \lesssim M_i$.
In this regime we can study the leptogenesis in the so-called one-flavour approximation.
The CP-violating asymmetry, summed over all flavours,
can be expressed in the form (for hierarchical heavy neutrinos):
\be
\epsilon_j \cong -\frac{3 M_j}{16\pi v^2}\, \frac{
\sum _{\beta}  {\rm Im}\,(
m^{2}_\beta \, R_{j\beta }^2) }{\sum_i m_i\, \left|R_{i}\right|^2}
\label{asym}
\ee 
where the orthogonal complex matrix $R$ contains all the information
on high energy phases of the seesaw model and is given by \cite{R} 
\be
R=vM^{-1/2}_{\rm{diag}} \hat{Y}_\nu U m^{-1/2}_{\rm{diag}}~,
\ee
where $\hat{Y}_\nu$ is neutrino yukawa coupling in the basis
of diagonal right-handed neutrinos.
Apparently the low energy phases 
present in $U_{\rm{PMNS}}$ do not play any role in generating lepton asymmetry.
However, in our context, the neutrino yukawa coupling $Y_\nu$ is subject to
the condition (\ref{yukt}) and we obtain a trivial $R$:
\be
R=\left(
\bad
\pm1&0&0\\
0&\pm 1&0\\
0&0&\pm 1
\end{array}
\right) ~.
\ee
As an immediate consequence 
the lepton asymmetry $\epsilon_j$ vanishes (even including flavour effects). 
It is important to remind
that the vanishing asymmetry is not a consequence of
the preserved $G_{23}$ symmetry.
Our viewpoint is quite different from that pointed out by authors of \cite{mu-tau2},
indeed, in their models the vanishing  $\epsilon_{j}$ 
are tightly due to the $\nu_\mu-\nu_\tau$ exchange symmetry.

The previous analysis, however, is expected to be modified by 
higher order corrections suppressed by $\langle \Phi \rangle /\Lambda$.
When these corrections are accounted for, $U_{e3}$ is no longer vanishing 
and, at the same time, the matrix $R$ will slightly differ from identity.	
Then, barring accidental cancellations, a non-vanishing $\epsilon_{j}$ is
related to non-vanishing Dirac or/and Majorana phases appearing in $U_{e3}$ . 
This result is quite general and the dependence of generated lepton asymmetry from 
low energy phases is dictated by flavor symmetries.  
However, since the next-to-leading order (NLO) corrections
are not determined by flavour symmetries,
a direct bridge between the BAU and the element $U_{e3}$ cannot
be generally established. In the following 
we will illustrate a case of leptogenesis in the $A_4$ seesaw model proposed in \cite{A4Lin} in which
a relationship between the low energy phases and the generated BAU arises naturally.

{\it{C.~A constrained seesaw model for TBM.}}~
So far the discussion was completely general. Now we
will focus on a special case of (\ref{allproposal}) 
with $\gamma= \pi/4$, $c=d$ and $a=-2d$ which corresponds to
a possible realization of TBM pattern.
Here we consider the seesaw model for TBM based on 
the flavour symmetry $A_4 \times Z_3 \times Z_4$ \cite{A4Lin}.
We recall that the discrete group $A_4$ is the group of even permutations of 4 objects 
and has one triplet and three singlet ($1$, $1'$, $1''$) representations. 

The lepton and EW Higgs content, together with their transformation properties 
under the flavour group, is displayed in Table~\ref{transform}.
The flavour symmetry breaking sector consists 
of the scalar fields neutral under the SM gauge group: $(\varphi_T, \xi')$ for
charged leptons, $(\varphi_S,\xi, \zeta)$ for neutrinos.
The additional $Z_3 \times Z_4$ discrete factor is needed in order to reproduce 
the desired alignment of scalar fields and 
at the same time allows a hierarchy between 
VEVs of scalars in different sectors as we will see in a moment.

For the charged lepton sector we choose $(\varphi_T, \xi') \sim (3,1')$ of $A_4$.
They are all invariant under $Z_3$ and carry a charge $i$ under $Z_4$.
It is not difficult to obtain a stable alignment \cite{A4Lin} of $(\varphi_T, \xi')$ as follows:
\be
\langle \varphi_T\rangle = (0,v_T,0)~,~~ \langle \xi' \rangle = u'~.
\ee
This structure of vacua automatically reproduces  
diagonal and hierarchical charged lepton masses through the following 
lagrangian:
\bea
-{\cal L}_e &=& \alpha_1 \tau^c (l \varphi_T) h_d/\Lambda \nn \\
&+& \beta_1 \mu^c \xi' (l \varphi_T)''h_d/\Lambda^2 +\beta_2 \mu^c (l \varphi^2_T) h_d/\Lambda^2  \nn \\
&+& \gamma_1 e^c (\xi')^2 (l \varphi_T)' h_d/\Lambda^3+
\gamma_2 e^c \xi' (l \varphi^2_T)'' h_d/\Lambda^3 \nn \\
&+& \gamma_3 e^c (l \varphi^3_T) h_d/\Lambda^3 + h.c. + \cdots  \nn
\label{wlplus}
\eea
After EW and flavour symmetry breakings one obtains:
\bea
m_l=\left(
\bad
\sim \varphi^3_T/\Lambda^3&0&0\\
0&\sim \varphi^2_T/\Lambda^2&0\\
0&0&\sim \varphi_T/\Lambda
\end{array}
\right) v_d~. \label{hierarchy}
\eea
The required hierarchy among 
$m_e$, $m_\mu$ and $m_\tau$ is approximately described
providing $$ \lambda^3_c \lesssim v_T/\Lambda \sim u'/ \Lambda \lesssim \lambda^2_c,$$ 
being $\lambda_c$ the Cabibbo angle.

\begin{table}
\caption{The transformation properties of leptons and
EW Higgs doublets under $A_4 \times Z_3 \times Z_4$.}
\begin{ruledtabular}
\begin{tabular}{|c||c|c|c|c|c|c|c|}
{\tt Field}& l & $e^c$ & $\mu^c$ & $\tau^c$ & $\nu^c$ & $h_u$ & $h_d$ \\
\hline
$A_4$ & $3$ & $1$ & $1$ & $1$ & $3$ & $1$ &$1$ \\
\hline
$Z_3$ & $1$ & $1$ & $1$ & $1$ &  $\omega$ & $1$ & $1$ \\
\hline
$Z_4$ &$1$ & $-1$ & $-i$ & $1$ & $1$ & $1$ & $-i$
\end{tabular}
\end{ruledtabular}
\label{transform}
\end{table}

The neutrino sector is given by the seesaw lagrangian with
3 heavy right-handed neutrinos $\nu^c_i$:
\bea
-{\cal L}_\nu = y (\nu^c l) \zeta h^u/\Lambda + 
x_a \xi (\nu^c \nu^c)+x_b (\varphi_S \nu^c \nu^c)+h.c. \nn
\label{seesaw}
\eea
where $(\varphi_S, \xi, \zeta) \sim (3,1,1)$ under $A_4$ and
have only $Z_3$ charge: $(\omega, \omega, \omega^2)$.
We find that the minimization of the scalar potential 
leads to the following VEVs:
\be
\langle \varphi_S\rangle = (v_S,v_S,v_S)~,~~ \langle \xi \rangle = u~,~~ \langle \zeta \rangle = v~.
\ee
The combinations $\zeta \xi$ and $\zeta \varphi_S$ are invariant under the
abelian part of the flavour group and can affect the charged lepton sector as corrections
at the next-to-next-to-leading order (NNLO). These corrections are
however suppressed by $1/\Lambda^2$ and we have a certain freedom to choose
flavour symmetry breaking scale in the neutrino sector
without destroying the hierarchical structure obtained
 for charged leptons (\ref{hierarchy}).
We will assume 
\be
 u/\Lambda \sim v_S/\Lambda \sim v/\Lambda \sim \lambda_c \div \lambda^2_c.
 \label{vevcond2}
 \ee
Differently from conventional models \cite{TBM} aimed to explain TBM,
this possibility can also describe a relatively large value of $\theta_{13}$ \cite{A4Lin}.
The leading contributions to the Dirac and Majorana masses are
\bea
m^D_0=\left(
\begin{array}{ccc}
1& 0& 0\\
0& 0& 1\\
0& 1& 0
\end{array}
\right) \frac {yvv_u} {\Lambda}, M=\left(
\begin{array}{ccc}
b+2 d& -d& -d\\
-d& 2d& b-d\\
-d& b-d& 2 d
\end{array}
\right)u\nn
\label{mnu0}
\eea
where
$b \equiv x_a$ and $d \equiv x_bv_S/u$.
As the general situation analyzed in the beginning, the leading order
mixing matrix, corresponding to TB mixing in this case, 
diagonalizes the right-handed neutrino mass matrix by (\ref{U0}). 
More precisely, $U_0=U_{\text{TB}} \Omega$ 
where  
\be
U_{\text{TB}}=\left(
\begin{array}{ccc}
\sqrt{2/3}& 1/\sqrt{3}& 0\\
-1/\sqrt{6}& 1/\sqrt{3}& -1/\sqrt{2}\\
-1/\sqrt{6}& 1/\sqrt{3}& +1/\sqrt{2}
\end{array}
\right)~,
\label{TB}
\ee
$\Omega=\rm{diag} \{ e^{i \phi_1/2},e^{i \phi_2/2}, i e^{i \phi_3/2} \}$ and
$\phi_1,\phi_2,\phi_3$ are respectively phases of $b+3d$, $b$, $b-3d$.
Moreover, the flavour symmetry automatically leads to a Dirac mass of the form 
(\ref{allproposal}), then the condition (\ref{yukt}) is fulfilled.
The physical masses of $\nu^c_i$ are given by $M_1=|b+3d|$, $M_2=|b|$ and $M_3=|b-3d|$
and those of light neutrinos are $m_i =  \left| y v_u v \right|^2 / (\Lambda^2 M_i) 
\sim y^2 \lambda^4_c v^2_u /M_i \div y^2 \lambda^2_c v^2_u /M_i$ .

In this model the only relevant NLO
corrections to the lagrangian ${\cal L}_e + {\cal L}_\nu$ appear in the Dirac mass
and they are of type $(\nu^c l \varphi \varphi) h^u/\Lambda^2$
 with $\varphi \in \{ \varphi_S, \xi \}$.
The correction to the Dirac mass  $\delta m^D$ that breaks the condition imposed by
 (\ref{yukt}) has the following form:
\bea
\delta m^D=\left(
\begin{array}{ccc}
0 & y_1+y_3& y_2-y_3\\
y_1-y_3& y_2& y_3\\
y_2+y_3& -y_3& y_1
\end{array}
\right) v_u \frac {v_S^2} {\Lambda^2}~, \nn
\eea
where $y_i$ are generally complex numbers of order $1$.
Including $\delta m^D$, the TB mixing should receive
a small correction according to $U_{\rm{PMNS}}=U_0 \delta U$.
The correction to leading mixing matrix $\delta U$ can be calculated perturbatively
and we find that, due to the special structure of $\delta m^D$,
only the $(13)$ and $(31)$ elements of $\delta U$ survive.
Moreover we should expect that $(\delta U)_{13} = -(\delta U)_{31} \sim \lambda_c \div \lambda^2_c$ 
depending on the scale of VEVs in (\ref{vevcond2}).
As a consequence, 
the tri-bimaximal prediction for $\theta_{12}$ remains unchanged 
at the first order in $(\delta U)_{13}$ 
and $(\delta U)_{13}$ simultaneously induces a departure of $\theta_{13}$ 
and of $\theta_{23}-\pi /4$ from zero.
Then we can derive the following sum-rule:
\be
\sin ^2 \theta_{23} = 1/2 (1+ \sqrt{2} \cos \delta \sin \theta_{13})+ O(\theta^2_{13})
\label{pred1}
\ee
where $\delta$ is the CP-violating Dirac phase in the standard parametrization
of $U_{\rm{PMNS}}$ . 

{\it{D.~Unflavoured Leptogenesis from low energy phases.}}~ 

As an illustration we will discuss a case of unflavoured leptogenesis 
{\footnote{For some other approaches concerning implication of leptogenesis 
in $A_4$ models see, for example, \cite{TBlepto}.}}
which mainly depends on low energy phases 
and its implication on reactor angle $\theta_{13}$.
We only consider the case of normal hierarchy for light neutrinos.
In this case, the right-handed neutrinos are
naturally hierarchical according to $M_3 \ll M_2 \approx 1/2 M_1$.
We can simply estimate the natural mass range of the lightest right-handed neutrino
$\nu^c_3$ by taking neutrino mass scale as $\sqrt{\Delta m^2_{\rm{atm}}}
\sim 0.05 \,\rm{eV}$
and the scale of $m^D$ as $v_u \lambda_c^4 \div v_u \lambda_c^2$ with $ v_u = 174 \,\rm{GeV}$
obtaining $M_3 \sim 10^{12} \,\rm{GeV} \div 3 \times 10^{13} \,\rm{GeV}$.

As the general analysis made in the beginning, leptogenesis
does work in this class of models only taking into account NLO corrections.
In the model presented in the previous chapter, although these corrections
are quite simple, their contribution to CP-violating asymmetries 
does not depend only on $(\delta U)_{13}$ but also on $\delta m^D$
itself. Indeed the NLO off-diagonal corrections to $R$ are perturbatively given by:
\bea
R &=& m^{1/2}_{\rm{diag}}~U^\dagger_0m^D_0 U_0~\delta U ~m^{-1/2}_{\rm{diag}} + \nn \\
&+& m^{1/2}_{\rm{diag}}~U^\dagger_0 \delta m^D U_0 ~m^{-1/2}_{\rm{diag}}~.
\label{R-new}
\eea
Since $U^\dagger_0m^D_0 U_0 = {\rm{diag}(1,1,-1)}$ the first row in (\ref{R-new})
is directly related with $(\delta U)_{13}$. However, the presence of  $\delta m^D$
generally destroys a hopeful alignment between $R$ and  $\delta U$.
This means that, without further considerations, 
a direct bridge between the BAU and the element $U_{e3}$ cannot
be established.

Now we observe that for normally ordered hierarchical light neutrinos,
it is suitable to consider the limit $m_1 \ll m_3$.
In this case it is not difficult to see that $(U^\dagger_0 \delta m^D U_0)_{31}
\simeq (\delta U)_{13}$. With this approximation, from (\ref{R-new})
one immediately finds that 
$$R_{31}=2\sqrt{m_3/m_1} (\delta U)_{13}~.$$
This is a very nice feature of this model with hierarchical neutrino spectrum because
a same small rotation matrix $\delta U$ is responsible both for corrections 
to TBM and in generating off-diagonal elements of R
responsible for leptogenesis. 

Now, it is convenient to parametrize $(\delta U)_{13}$ in terms of low energy physical quantities 
$U_{e3} = e^{-i\delta} \sin \theta_{13} $ and 
$\phi_{13}=(\phi_1 - \phi_3)/2$:
\be
(\delta U)_{13} = \sqrt{3/2} \, e^{-i \phi_{13}} U_{e3}~.
\ee
From (\ref{asym}) we obtain a lepton asymmetry for the decay
 of $\nu^c_3$ which explicitly depends on the low energy Dirac phase $\delta$
 and the physical Majorana phase $\phi_{13}$:
 \bea 
\epsilon_3 &=& -\frac{3 M_3 m_1}{4\pi v_u^2}\,  {\rm Im}\, [(\delta U)^2_{13}]  \\
&=&
\frac{9 M_3 m_1}{8\pi v_u^2}\, \sin (2\delta+ 2\phi_{13}) \sin^2 \theta_{13} \nn
\label{asym2} ~.
\eea
We are in the strong wash-out regime since
$$\tilde{m}_3=\sum_\alpha m_\alpha |R_{3 \alpha}|^2  \approx m_3 \approx 
(\Delta m^2_{\rm{atm}})^{1/2}~.$$ 
In this regime, the final lepton asymmetry can be approximately reproduced by \cite{lepto}:
$$ Y_L \cong 0.3 \frac{\epsilon_3}{g_*} 
\left (\frac{0.55 \times 10^{-3} \rm{eV}}{ \tilde{m}_3} \right)^{1.16} ~.$$
Using the observed central value of $Y^{\rm{obs}}_B =8.6 \times 10^{-11}$ \cite{BAU}
we obtain the following lower bound on $\sin^2 \theta_{13}$:
\be
\sin^2 \theta_{13} \gtrsim 0.005  \times \frac{3 \times 10^{13} \,\rm{GeV}}{M_3}
\ee
where we have used $g_* = 217/2$ and $m_1 =
(\Delta m^2_{\rm{sun}})^{1/2} \sim 0.01\, \rm{eV}$.
Since $M_3$ cannot be too much larger than $10^{13} \,\rm{GeV}$, 
a neutrino spectrum of normal hierarchy in this model favors  
a value of $\sin^2 \theta_{13}$ larger than $0.005$.
 
In summery, we have considered a general class of seesaw models
in which leptogenesis can be related with low energy CP violating phases.
The flavour symmetry which leads to a vanishing $U_{e3}$ at leading order
plays a central role. We gave an explicit model realization of this correlation
which also explains the TBM. The same approach can also be applied 
to many other flavour models \cite{parity, mu-tau, LLL, BI, TBM, hierarchy}
and it demonstrates that, even in flavoured leptogenesis, a connection between
low and high energy CP violation can be established naturally.

This work has been partly
supported by the European Commission under contract MRTN-CT-2006-035505

\end{document}